# Lubrication-Induced Newtonianization Enables Passive Transport of Non-Newtonian materials

Arvind Arun Dev[1,2*], Paszkal Papp[3], Thomas M. Hermans[3], Bernard Doudin[1]
1 IPCMS UMR 7504, Université de Strasbourg, 23 Rue du Loess, Strasbourg 67034, France
2 Laboratoire Colloïdes et Matériaux Divisés, CBI, ESPCI Paris, Université PSL, CNRS, 75005 Paris, France.
3 IMDEA Nanociencia, C/ Faraday 9, 28049 Madrid, Spain
*Corresponding author. Email: arvind.dev@espci.fr

## Abstract

Non-Newtonian flows are typically governed by intrinsic bulk rheology, which imposes strong constraints on transport through confined geometries. Here, we show that stable boundary lubrication can fundamentally alter this behavior by localizing shear within a thin, low-viscosity interfacial layer. As a result, the nonlinear rheological response of a broad class of complex materials—including yield-stress, shear-dependent, and thixotropic materials—are strongly suppressed during flow. Using analytical solutions of Stokes flow and numerical simulations, we demonstrate that lubrication-induced shear localization leads to an apparent Newtonianization of transport, in which the macroscopic flow response becomes primarily controlled by the lubricating layer and geometric confinement rather than the intrinsic material properties. In this regime, materials that would otherwise require large pressure gradients can be transported at substantially lower driving forces. Notably, this boundary-dominated transport enables gravity-driven passive flow with orders-of-magnitude enhancement in throughput compared to rigid-wall conduits. These results establish lubrication as a powerful mechanism to tune and simplify complex fluid transport, with implications for biological systems, soft and jammed materials, and energy-efficient fluid .

## Introduction

Complex materials exhibiting Non-Newtonian rheology, with viscosity changing under shear stress $\tau$ , span a vast spectrum of biological(*1*–*3*), industrial(*4*), pharmaceutical(*5*), defense(*6*) and environmental systems(*7*). Gels, pastes, many polymers, blood and biological tissues are common examples of their complicated rheological behavior (see Table 1). The non-linearity results from shear-induced structural (re)arrangements of constituents of materials(*8*, *9*), non-local rheology(*10*–*12*), elastoplastic behavior(*13*), and their feedback to the applied stress (Fig. 1a)). These microstructural events impact macroscopic properties. Some typical behaviors include shear-thinning and shear-thickening (*14*, *15*), where the material's resistance to flow decreases and increases with applied shear rate $\dot{\gamma}$ respectively. Another class is made of yield stress materials(*16*, *17*), where the material does not flow (no shear) below a threshold stress value. Furthermore, time-dependent yielding which is termed as thixotropic behavior(*18*, *19*), includes breaking and formation of material's internal structure over time leading to time dependent viscosity. Each class requires a specific phenomenological constitutive model relating stress and strain rate (Table 1). Unlike Newtonian materials, no single model can encompass all of these bulk rheological behaviors, complicating both theoretical predictions and experimental reproducibility. The universality of these materials extends further into sustainable energy applications,

| Table 1: Observed Rheology of Non-Newtonian materials and their constitutive model | | | |
|---|---|---|---|
| Observed rheology | System modelled | Constitutive model (name & equation) | Reference |
| Yield-stress & Shear-thinning | Colloidal gels, foams, microgels, drilling fluids, whole blood, biological tissues. | Herschel–Bulkley $\tau = \tau_y + k\dot{\gamma}^n\ , n < 1$ | (*1*, *20*–*22*) |
| Yield-stress & Shear-thickening | Bauxite Residue (Red Mud), Silica/PEG suspensions, Magneto-rheological fluids | Herschel–Bulkley $\tau = \tau_y + k\dot{\gamma}^n\ , n > 1$ | (*6*, *8*, *17*) |
| Shear-thinning (Power Law) | dilute polymer solutions, xanthan gum, wall paints. | Ostwald-de Waele $\tau = k\dot{\gamma}^n\ , n < 1$ | (*23*, *24*) |
| Shear-thickening (Power Law) | Cornstarch/Water (Oobleck), Ethylene Glycol, TiO2 suspensions. | Ostwald-de Waele $\tau = k\dot{\gamma}^n\ , n > 1$ | (*15*, *24*) |
| Time-dependent (Thixotropic) | Carbon black solutions, drilling cuttings fluids, synovial joint fluids. | $\frac{\partial S}{\partial t} + u \cdot \nabla S = k_b(1 - S) - k_d S\dot{\gamma}$ $\eta(S) = \eta_\infty + (\eta_0 - \eta_\infty)S$ $\eta_0 = 5\ \mathrm{Pa \cdot s}\ , \eta_\infty = 0.5\ \mathrm{Pa \cdot s}$ | (*18*, *25*, *26*) |

where the high viscosity of carbon-dense solvents in carbon capture(*4*), phase-change slurries in energy storage(*27*, *28*) and dense suspensions in flow battery govern the parasitic energy loss (*29*, *30*) and limits the scalability. It becomes particularly critical in confined environments, where jammed materials like concentrated suspensions(*16*), emulsions, drug formulations(*31*), and cell ensembles(*1*) present highly non-linear complex rheological characteristics. It often leads to prohibitive shear rates and structural degradation, ultimately hampering functionality, especially when flow circuits are reduced in size. Despite their wide prevalence, the transport of non-linear materials through rigid confined geometries suffers from large pressure drops and pumping infrastructure of disproportionate size. Consequently, a universal challenge remains: how to mitigate these non-linearities during transport, without altering the materials? A non-intrusive solution is to rely on lubricated transport(*32*), which results in low dissipative states(*33*–*36*).

Conventional techniques with passive lubricant-infused surfaces significantly mitigate viscous dissipation(*35*) but are limited by capillary retention, require surface engineering, surface micro structuring, and are prone to shear induced lubricant depletion under large throughput(*37*). Instabilities and non-standard geometries have further restricted the studies from exploring the rheological regimes of lubrication mediated transport.
Here we propose to use a ferrofluid lubricant (Fig 1b), motivated by our recent experimental studies on stabilizing such lubricant under properly designed magnetic force field gradient (*38*–*41*). With this design strategy, the shear is localized within the thin and low-viscosity lubricant film. We detail how the boundary-controlled rheology mitigates the energy-intensive intrinsic non-linear signature of the transported material. The key outcome is that the macroscopic transport response undergoes an "apparent Newtonianization" where shear thinning and shear thickening materials behave like identical Newtonian liquids, providing universality, predictability and keeping the material integrity when flowing. Furthermore, we quantify how the material flow throughput can be greatly enhanced. We detail the case of how the simple gravitational pressure gradient overcomes the viscous shear and results in flow rates which are orders of magnitude higher than the non-lubricated case.

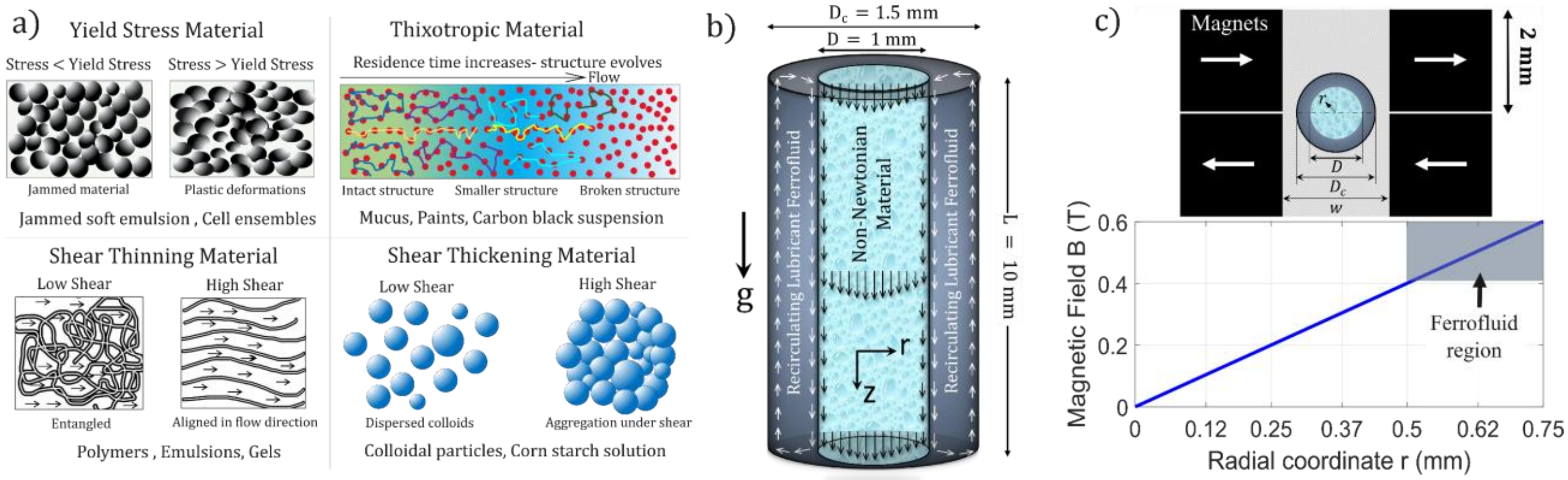

Figure 1 Non-Newtonian materials and transport. a) Different non-Newtonian materials, mechanism of non-linear behavior and their qualitative behavior under shear. Schematics motivated by (9, 11, 42, 43). b) Lubricated transport with recirculating lubricant (ferrofluid) in the annulus region and non-Newtonian material in the core of diameter D. $\mathrm{D_c}$ is cavity diameter. c) Magnetic design to form core -annulus flow structure. Four magnets of 2 mm× 2 mm × L mm, with remnant magnetization $\mathrm{M_r}$ = 1.26 T are arranged in quadrupolar geometric design. The arrows (white) show the direction of magnetization of magnets. The distance between pair is w = 2 mm. Magnetic field intensity B varies linearly (∝ r) in this design(38). The annulus region (ferrofluid) is in high magnetic field (see shaded).

We employ low Reynolds number Stokes flow and numerical simulations using ANSYS CFX to get insight into these regimes of lubrication, making our findings relevant for a broad class of materials.

**Material and setup:** One of the classic phenomenological constitutive models used for a non-Newtonian material is the Herschel-Bulkley model(*22*) with $\tau = \tau_y + k\,\dot{\gamma}^n$. In this framework, $\tau_y$ is the yield stress, the threshold below which the material exhibits no shear and behaves as a solid-like material, and $\dot{\gamma}$ is the shear rate. The parameters $k$ and $n$ denote the consistency index and the flow behavior index, respectively. The effective viscosity is subsequently derived as: $\eta_{eff} = \frac{\tau_y}{\dot{\gamma}} + k\,\dot{\gamma}^{n-1}$. The $n$ categorizes material response such that: for $n = 1$ ($\tau_y = 0$), we recover the Newtonian characteristics; for $n < 1$, shear-thinning material, where $\eta_{eff}$ decreases with increasing $\dot{\gamma}$, and for $n > 1$, shear-thickening material, where $\eta_{eff}$ increases with increasing $\dot{\gamma}$. Table 1 shows other relevant constitutive models and materials governed by them.

The typical flow response would change dramatically when a lubricant is present at the wall (Fig. 1b) to form a core-annular flow structure. In the core of diameter $D = 1$ mm, flows the non-Newtonian material and the lubricant is a magnetic liquid (ferrofluid) placed near a wall in a cavity of diameter $D_c = 1.5$ mm forming the annulus region. Ferrofluids are mildly shear thinning; decreasing $\eta_f$ on increasing $\dot{\gamma}_f$, or Newtonian at low shear with constant $\eta_f$ (*38*, *44*). We focus on Newtonian ferrofluid to concentrate on the scaling of non-linearity of the transported material. The effect of non-linear ferrofluid is discussed in SI S1.

For lubricated transport, a quadrupolar magnetic field is used to spatially confine a ferrofluid, creating a high-field region near the magnets and a field minimum at the centre (Fig. 1c). The cross section of the flow arrangement shows magnetic design with bar magnets of size 2×2 mm arranged in a quadrupolar arrangement with magnet pair separated by a distance $w$ = 2 mm (*38*). Under this geometric arrangement the magnetic field can be approximated as invariant in azimuthal direction and increasing linearly in the radial direction. The magnetic field, given by (*38*) $B = 4\mu_0 M_r r/\pi w$ is zero at the center of the flow channel ($r = 0$) and maximum of 0.6 T at the cavity wall ($r = D_c/2$) (*38*). Here $M_r = 1.26\,T$ is the remnant magnetization of the magnets. This field topology stabilizes a ferrofluid annulus while inducing the formation of an inner anti-tube of the transported liquid along the low-field core. The annulus region with ferrofluid is under high magnetic field (shaded region in Fig. 1c)) and saturates magnetically with a saturation magnetization $M_s$. Generally, $M_s$ ranges from 10-100 mT with saturation field $B$ = 0.2 T (*44*). The persistence of this core-annulus structure is governed by the balance between magnetic confinement and competing mechanical stresses. The magnetic energy density $E_m$ (magnetic pressure) that stabilizes the structuure scales as $E_m \sim \mu_0 M_s M_r d/w$ where $\mu_0$ is the magnetic permeability (*38*). During flow of the non-Newtonian liquid, viscous shear stress acts to deform the interface and for large viscosity ratio ($\eta_r \gg 1$) scales as $\tau \sim 32Qk\dot{\gamma}^{n-1}/\eta_r D^3$. The relative strength is given by Mason number(*44*) $Mn \sim \tau/E_m$. in our system, $Mn < 10^{-3}$, indicates that viscous stresses are insufficient to deform the core-annulus interface. Capillary stresses tend to minimize interfacial area and can destabilize confined liquid–liquid structures. The ratio of interfacial energy density $E_s$ to magnetic energy density is given by the magnetic Bond number(*40*), $Bo_m = \frac{E_s}{E_m} = \frac{\pi w \sigma}{4\mu_0\ Ms\ Mr}$, which for the present system, $Bo_m << 10^{-3}$, demonstrating that interfacial tension cannot destabilize the ferrofluid interface or induce hydrodynamic instabilities. Moreover, gravitational effects are also dominated by magnetic forces, since the maximum gravitational energy density for ferrofluid ($E_g = \rho_f g L$) and the corresponding magneto-gravitational Bond number is $Bo_{mg} = \frac{E_g}{E_m} = \frac{2\rho_f g h_{max}}{\mu_0 M_s M_r} << 10^{-2}$, confirming that magnetic confinement is sufficient to hold the ferrofluid against gravity. (see SI S2 for details for scaling laws). Thus, in this configuration, the ferrofluid annulus, with a viscosity of 0.01 $\mathrm{Pa \cdot s}$ and a cavity diameter of $D_c = 1.5$ mm, effectively functions as a non-deformable wall that lubricates the flow of the inner non-Newtonian material.

**Model of time-independent non-Newtonian materials**

We consider an axially symmetric cylindrical core-annulus flow (Fig. 1b and Fig. 1c). The non-Newtonian material is in the core (diameter $D$) encapsulated by a magnetic lubricant in the annular thickness, $\left(\frac{D_c - D}{2}\right)$. We model the Stokes flow in the limit of small Reynolds number ($Re \ll 1$), therefore, the flow is given by

$$-\frac{\partial P}{\partial z} = \left(\frac{1}{r}\frac{\partial}{\partial r}(r\tau)\right) \quad (1)$$

Where $\partial P/\partial z$ is the pressure gradient and $\tau$ is the shear stress. Integrating Eq.1 we get

$$\tau = -\frac{r}{2}\frac{\partial P}{\partial z} + \frac{A}{r}, \quad (2)$$

and in order for the stress to be finite at $r = 0$, we have $A = 0$ and hence, $\tau = -\frac{r}{2}\frac{\partial P}{\partial z}$. The non-Newtonian material is characterized by Herschell-Bulkley law as

$$\tau = \tau_y + k\,\dot{\gamma}^n. \quad (3)$$

Using Eq. 3, we get $\dot{\gamma}$ as

$$\dot{\gamma} = -\frac{\partial u}{\partial r} = \left[-\frac{r}{2k}\frac{\partial P}{\partial z} - \frac{\tau_y}{k}\right]^{\frac{1}{n}} \quad (4)$$

Where $u(r)$ is the axial velocity of flow. For a confined flow (here Poiseuille type flow), $\tau \propto r$. Assuming that for $r \leq R_0$ we have $\tau \leq \tau_y$, hence the material does not yield for $r < R_0$ and moves as a solid plug (with no shear) of constant velocity $u_0$. Hence at $r = R_0$, $\dot{\gamma} = -\frac{\partial u}{\partial r} = 0$, and $\tau = \tau_y$. Using this we get

$$\tau_y = -\frac{R_0}{2}\frac{\partial P}{\partial z} \quad (5)$$

Using Eq. 4 and Eq. 5 we get

$$\frac{\partial u}{\partial r} = -\left(-\frac{1}{2k}\frac{\partial P}{\partial z}\right)^{\frac{1}{n}} (r - R_0)^{1/n} \quad (6)$$

Integrating, we get $u(r)$, the velocity profile.

$$u(r) = \frac{-1}{\frac{1}{n}+1}\left(-\frac{1}{2k}\frac{\partial P}{\partial z}\right)^{\frac{1}{n}} (r - R_0)^{\frac{1}{n}+1} + B \quad (7)$$

with $B$ as integration constant. For the lubricant (ferrofluid), we consider the ferrofluid as a Newtonian liquid with $n_f = 1$, and the viscosity of ferrofluid ($\eta_f$) as constant in the magnetically saturated scenario(*39*). The subscript $f$ is used to denote variables associated with ferrofluid. Using Eq.1 with $-\eta_f\frac{\partial u_f}{\partial r} = \tau_f$ and integrating twice, we get

$$u_f(r) = \frac{r^2}{4\eta_f}\frac{\partial P_f}{\partial z} + C\ln(r) + E \quad (8)$$

To find the constants $B$, $C$, and E we have following boundary conditions. Wall velocity is zero at the cavity wall, no-slip condition

$$r = \frac{D_c}{2}, u_f = 0 \quad (9)$$

The ferrofluid volume is conserved and hence there is no net flow across any cross section.

$$Q_f = 2\pi\int_R^{D_c/2} r u_f\, dr = 0 \quad (10)$$

Where $R = D/2$ is the radius of the core. The stress and velocity continuity at the ferrofluid-non-Newtonian material interface

$$r = R, u_f = u \quad (11)$$

$$r = R, -\eta_f\frac{\partial u_f}{\partial r} = \tau \quad (12)$$

Using Eq. 10 we get

$$E = -\frac{D_c^2}{16\eta_f}\frac{\partial P_f}{\partial z} - C\ln\left(\frac{D_c}{2}\right) \quad (13)$$

Using Eq. 11 we get

$$\frac{1}{16\eta_f}\frac{\partial P_f}{\partial z}R^4 a_1 + \frac{D}{2}R^2 a_2 + \frac{C}{2}R^2 a_3 = 0 \quad (14)$$

With $a_1 = \left(\frac{D_c}{2R}\right)^4 - 1$, $a_2 = \left(\frac{D_c}{2R}\right)^2 - 1$ and $a_3 = \left(\frac{D_c}{2R}\right)^2 \ln\left(\frac{D_c}{2}\right) - \frac{1}{2}\left(\frac{D_c}{2R}\right)^2 - \ln(R) + \frac{1}{2}$

Using Eq. 13 and Eq. 14, we get

$$C = \frac{\frac{-1}{16\eta_f}\frac{\partial P_f}{\partial z}R^2 a_4}{a_5},\ E = \frac{-1}{16\eta_f}\frac{\partial P_f}{\partial z}R^2 a_6 \quad (15)$$

With $a_4 = a_1 - \frac{a_2}{2}\left(\frac{D_c}{R}\right)^2$, $a_5 = \frac{a_3}{2} - \frac{a_2}{2}\ln\left(\frac{D_c}{2}\right)$ and $a_6 = \frac{a_4}{a_5}\ln\left(\frac{D_c}{2}\right) - \left(\frac{D_c}{R}\right)^2$

Using Eq. 15 and Eq. 14 in Eq. 8, we get the velocity profile in the ferrofluid as

$$u_f(r) = \frac{1}{4\eta_f}\frac{\partial P_f}{\partial z}\left[r^2 - \frac{a_4 R^2}{4a_5}\ln(r) + \frac{a_6 R^2}{4}\right] \quad (16)$$

Using Eq. 12 with Eq. 16, along with $\tau = -\frac{r}{2}\frac{\partial P}{\partial z}$ we get

$$\frac{\partial P_f}{\partial z} = \frac{\frac{\partial P}{\partial z}}{a_7} \quad (17)$$

with $a_7 = \left(1 - \frac{a_4}{8a_5}\right)$. Using Eq. 11, we get

$$B = \frac{a_8}{4\eta_f}\frac{\partial P}{\partial z} + \frac{1}{\frac{1}{n}+1}\left(-\frac{1}{2k}\frac{\partial P}{\partial z}\right)^{\frac{1}{n}}(R-R_0)^{\frac{1}{n}+1} \quad (18)$$

and finally, the velocity profile in the non-Newtonian material is given by

$$u(r) = \frac{1}{\frac{1}{n}+1}\left(-\frac{1}{2k}\frac{\partial P}{\partial z}\right)^{\frac{1}{n}}\left[(R-R_0)^{\frac{1}{n}+1} - (r - R_0)^{\frac{1}{n}+1}\right] + \frac{a_8}{4\eta_f}\frac{\partial P}{\partial z} \quad (19)$$

With $a_8 = \left[R^2 - \frac{a_4 R^2}{4a_5}\ln(R) + \frac{a_6 R^2}{4}\right]/a_7$

We also know that for $0 < r < R_0$, we have plug profile given by $u_0 = u(R_0)$

$$u_0 = \frac{1}{\frac{1}{n}+1}\left(-\frac{1}{2k}\frac{\partial P}{\partial z}\right)^{\frac{1}{n}}\left[(R-R_0)^{\frac{1}{n}+1}\right] + \frac{a_8}{4\eta_f}\frac{\partial P}{\partial z}$$
(20)

Using Eq. 19 and Eq. 20 and knowing that flow rate is given by $Q = \int_{R_0}^{R} u(r) 2\pi r\, dr + \int_0^{R_0} u_0 2\pi r\, dr$

The total flow rate $Q$ is given by

$$Q = \frac{\left(-\frac{1}{2k}\frac{\partial P}{\partial z}\right)^{\frac{1}{n}}2\pi}{\frac{1}{n}+1}\left(a_9 + \frac{R_0^2(R-R_0)^{\frac{1}{n}+1}}{2}\right) + \frac{\pi a_8}{4\eta_f}\frac{\partial P}{\partial z}(R^2) \quad (21)$$

with $a_9 = \left[\frac{(R^2-R_0^2)(R-R_0)^{\frac{1}{n}+1}}{2} - \frac{R(R-R_0)^{\frac{1}{n}+2}}{\frac{1}{n}+2} + \frac{(R-R_0)^{\frac{1}{n}+3}}{\left(\frac{1}{n}+2\right)\left(\frac{1}{n}+3\right)}\right]$

Eq. 21 can also be seen as

$$Q = Q_r + Q_{lub} \quad (22)$$

$Q_r$ corresponds to the flow rate through a rigid wall tube (non-lubricated) with zero wall velocity at radius *R*, while $Q_{lub}$ is the additional flow rate due to finite velocity at radius *R* induced by the lubricating layer:

$$Q_{lub} = \frac{\pi a_8}{4\eta_f}\frac{\partial P}{\partial z}(R^2) \quad (23)$$

**Model of time-dependent non-Newtonian materials**

In these materials, thixotropic systems, the material microstructure evolves temporally under constant stress, resulting in a time-dependent rheological response. This evolution is described through a structural parameter $S$, governed by competing structure build-up and shear-induced breakdown processes (Table 1). The structure parameter $(S)$ is modelled using computational fluid dynamics package ANSYS CFX by a transport equation

$$\frac{\partial S}{\partial t} + u \cdot \nabla S = k_b(1-S) - k_d S\dot{\gamma}, \quad (26)$$

where $u$ is the velocity and $S$ is the structure parameter. The structure parameter $S$ evolves such that $0 \le S \le 1$ where $S = 1$ and $S = 0$ indicates fully intact and completely broken material structure respectively. The intrinsic structural relaxation time $\tau_{\text{struct}} \sim \frac{1}{k_b} = 5\ s$ sets the timescale for recovery, while the breakdown timescale $(k_d\dot{\gamma})^{-1}$ controls shear-induced structure breaking ($k_d = 1$ in simulations), together defining the transient viscosity $\eta_{eff}(S)$,

$$\eta_{eff}(S) = \eta_\infty + (\eta_0 - \eta_\infty)S \quad (27)$$

with $\eta_0 = 5\ \mathrm{Pa \cdot s}$, and $\eta_\infty = 0.5\ \mathrm{Pa \cdot s}$ are viscosities at zero shear and maximum shear, respectively, making $S = f(t)$(*18*).

**Results and discussion**

**Power law materials:** A power law material behavior is a limiting case without yield stress $(\tau_y = 0)$ with $\tau = k\,\dot{\gamma}^n$ (shown in Table 1).

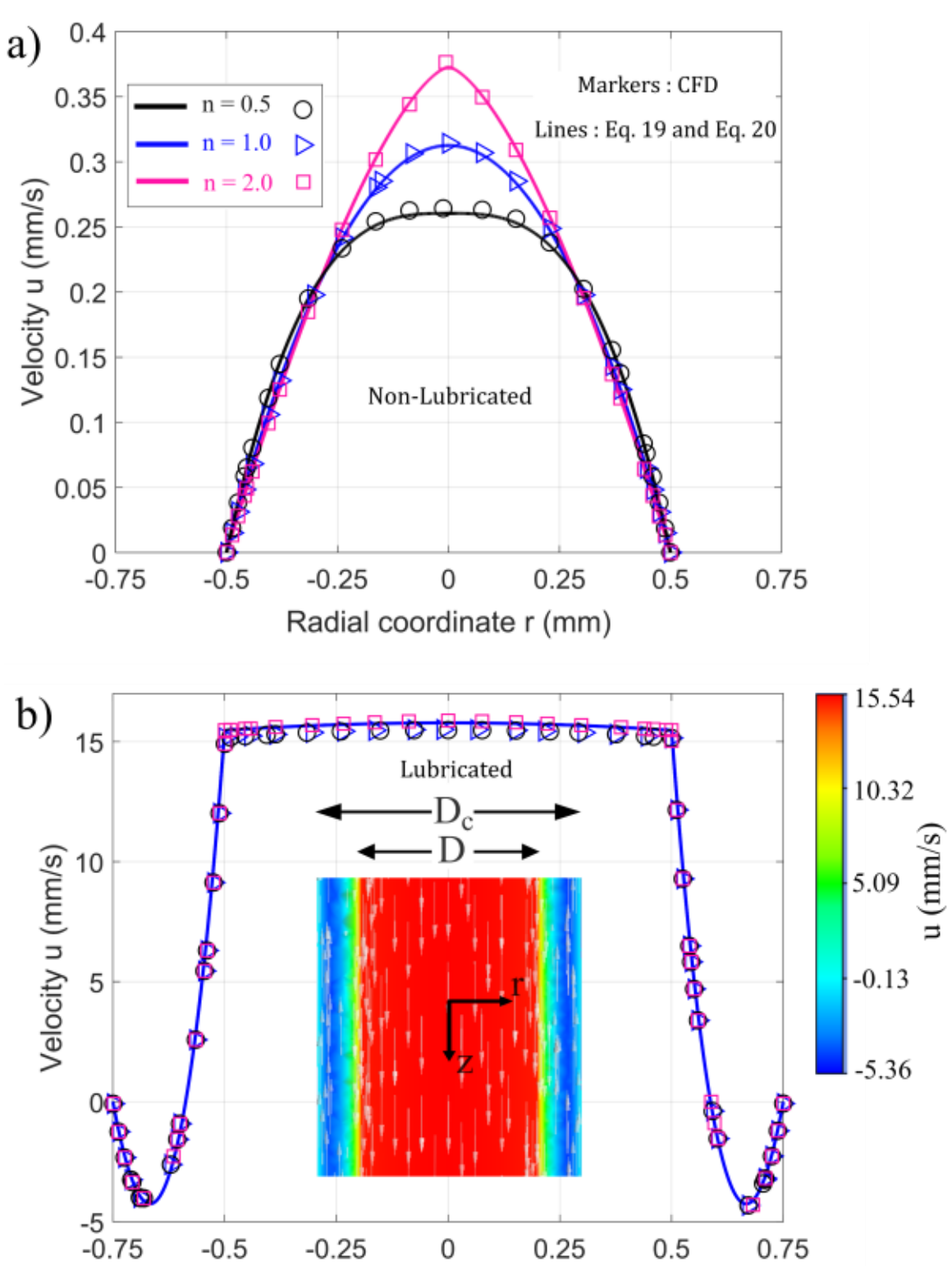

Figure 2 Flow of power law materials. a) and b) presents the velocity profile across the flow channel in fully developed flow for non-lubricated and lubricated confinement respectively. The pressure gradient and consistency index is fixed for n = 0.5, n = 1, n= 2 as $-10^4$ Pa/m and 2 Pa$\cdot$ $s^n$ respectively. Lines are Stokes solution using Eq.19 Eq.20 and Eq.16. Markers are CFD simulations (inset Fig. 2b).

Fig. 2a) and Fig.2b) show the flow through a non-lubricated and lubricated confinement respectively. The core diameter $D$, consistency index $k$ and gravitational pressure gradient is fixed at $1\,\mathrm{mm}$, $2$ Pa$\cdot$ $s^n$ and $-10^4\,\mathrm{Pa/m}$ respectively. For non-lubricated flow (Fig. 2a)) the velocity profile shows distinct variations. We compute flattened velocity profile for $n = 0.5$ and skewed for $n = 2$ when compared to Newtonian $n = 1$ (blue line). For lubricated flow with ferrofluid lubricant of viscosity $\eta_f = 0.01\,\mathrm{Pa}\cdot\mathrm{s}$ (Fig. 2b), the distinction between velocity profiles is difficult. The velocity profiles for $n = 0.5$ and $n = 2$ almost coincides with each other and also with a Newtonian liquid in lubricated flow with $n = 1$ (blue line). This is due to the shear localization in the ferrofluid layer. Under shear, the ferrofluid recirculates in the annulus region to conserve the volume, seen in inset. The core material flows with almost no shear and hence the change in the material property is suppressed. The satisfactory match between CFD (markers) and analytical model (Eq. 16, Eq. 19 and Eq. 20 with $R_0$= 0) further validates our CFD protocol (see SI S1 for CFD details).

Due to ultra-low shear ($\approx 0$), the flow rate increases by two orders of magnitude to $Q$ = 736.41 $\mu$L/min and $Q$ = 736.58 $\mu$L/min for $n = 0.5$ and $n = 2$ respectively compared to non-lubricated case with $Q_r = 7.37$ $\mu$l/min, 7.50 $\mu$l/min for $n = 0.5$, and $n = 2$ respectively. The flow rate $Q$ increases with applied pressure gradient.

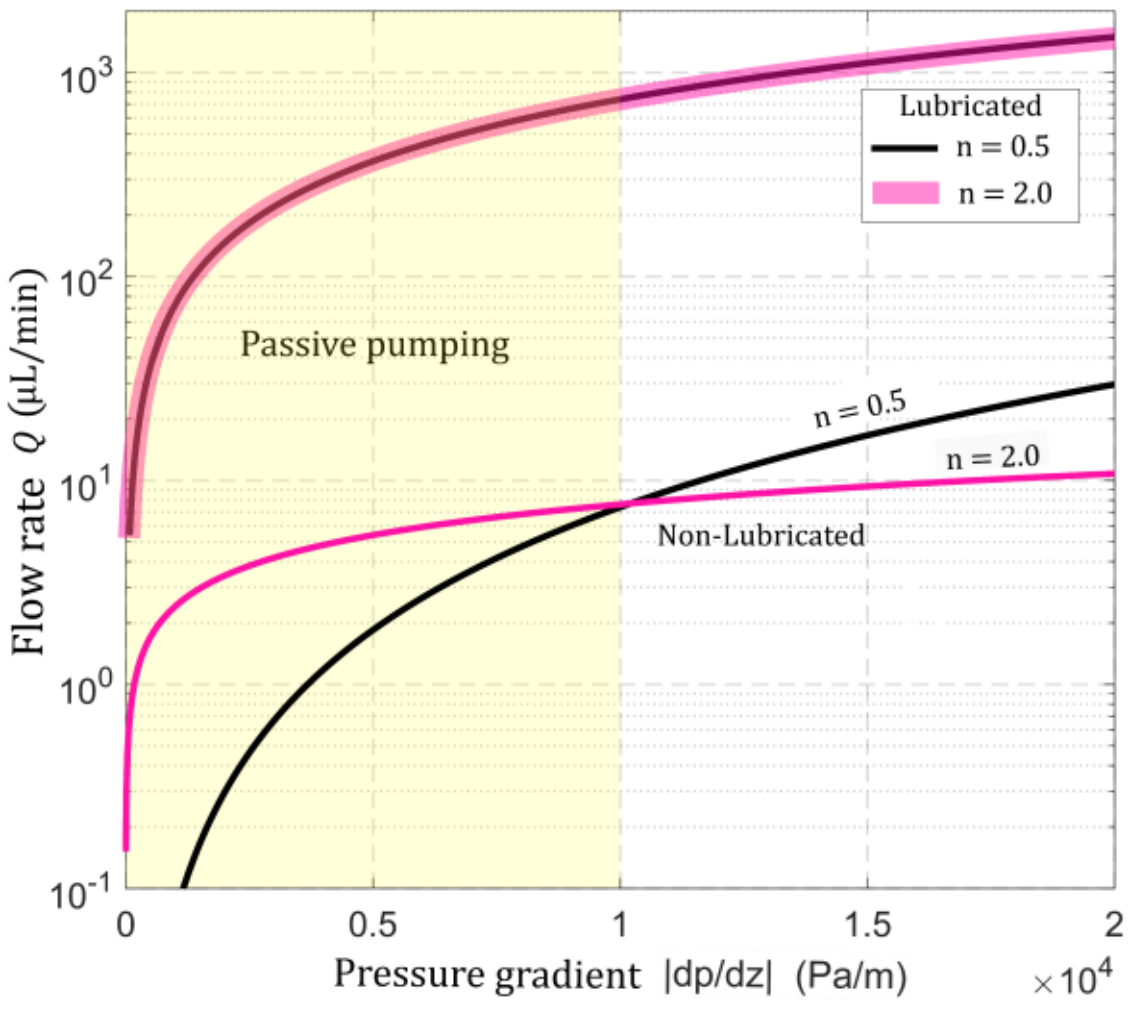

Figure 3 Flow of power law materials. a) and b) presents the flow curves relating Flow rate Q and pressure gradient for Rigid wall which is non-lubricated and lubricated wall flow for shear-thinning n = 0.5 and shear-thickening n = 2 materials respectively. For n = 2 in lubricated case, the line is thick to highlight the co-incident plots. The shaded region highlights the passive pumping region where the pressure gradient is below gravity.

Fig. 3 shows flow curves (flow rate vs pressure gradient) while flowing through a non-lubricated channel and a lubricated channel. For $n = 0.5$, and increasing pressure gradient, the flow rate $Q_r$ in a rigid (non-lubricated) tube of diameter 1 mm ranges from 1-10 $\mu l/\text{min}$, whereas for an identical pressure gradient, in a lubricated arrangement, the flow rate $Q = Q_r + Q_{lub}$ is up to 1 $\text{ml/min}$, here $Q_{lub}$ is the enhancement due to lubrication (Eq. 24 and Eq. 25). A two-order-of-magnitude enhancement in flow rate is observed along the flow curve. This high-throughput arrangement also holds true for a shear-thickening liquid, where we again observe a two-order-of-magnitude improvement in flow rate. In contrary to the non-lubricated case, the flow curves for lubricated case overlap for shear thinning and shear thickening materials. The shear thickening material is plotted as a band to highlight the overlapping. Furthermore, the shaded regions in Fig. 3 correspond to the region below the gravitational pressure gradient ($\rho g = -10^4$ Pa/m). Note that here the magnetic lubricant can be held in place by using permanent magnets(*38*), that recirculates in its position. Hence, it does not need continuous power input, and the non-linear material in the shaded region flows under gravity; hence, '*no mechanical pumping*'. A true passive pumping behavior with a large flow rate and ultra-low ($\approx 0$) active energy input can be realized. It is noteworthy that since the shear is mostly localized in the ferrofluid lubrication layer, and $Q_{lub} \gg Q_r$ , the total flow rate $Q \approx Q_{lub}$ for a fixed pressure gradient. Since $Q_{lub}$ depends only on the viscosity of the lubricant (Eq. 23), the overall flow behavior is governed by the lubricant viscosity and the channel geometry, rather than the non-Newtonian properties of the core material. Tuning the interfacial shear dynamics allows materials to flow without being strongly limited by their constitutive properties, making it possible to reach flow rates that standard rigid-wall systems cannot achieve. Importantly, this effect is observed even under passive pumping conditions (below the gravitational pressure gradient), offering a clear advantage over existing lubricant-based transport methods.

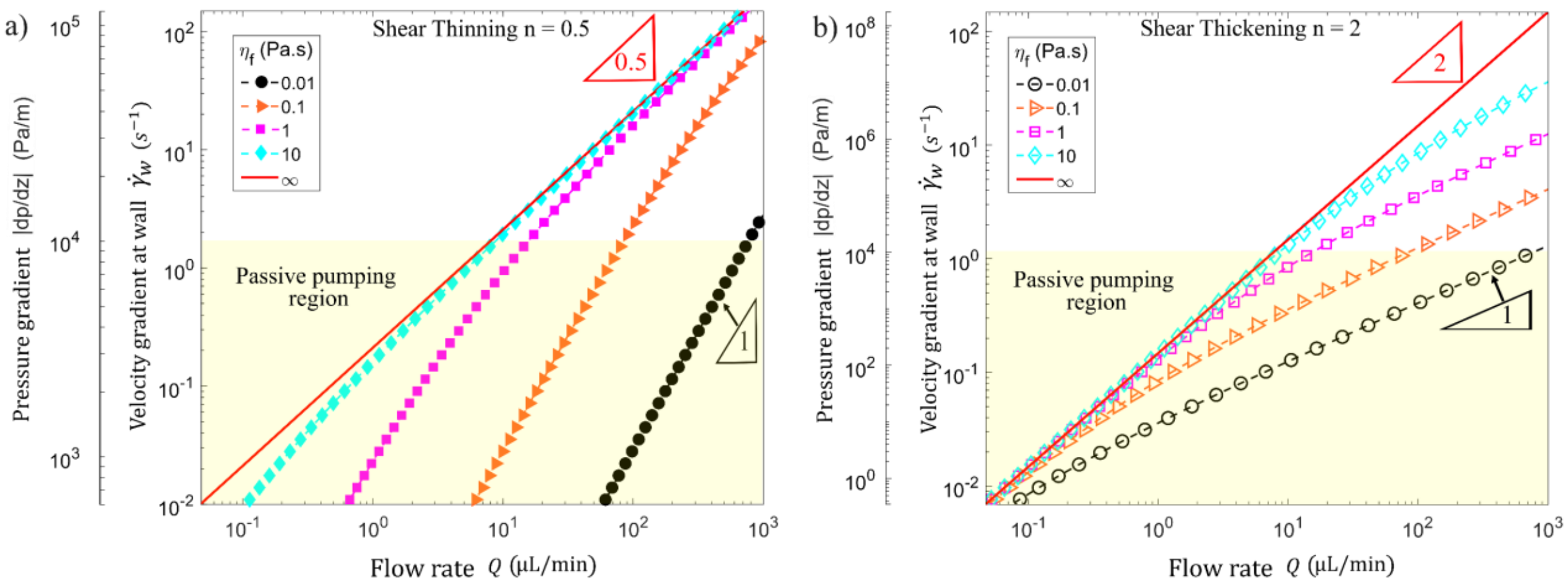

Figure 4 Flow rate (Q ) and maximum velocity gradient (maximum shear rate) $\dot{\gamma}_w$ (at wall) for shear thinning (Fig. 4a) and shear thickening (Fig. 4b) materials with different lubricant viscosity. Shaded region is where passive pumping can be realized. The slope of $\left|{}^{\partial P}/_{\partial z}\right|$ versus Q in log-log plot gives the power law index n of the material.

For a fixed $d$, $n$ and $k$, tuning the shear localization is possible by changing the lubricant viscosity. Fig. 4(a) and Fig 4(b) show the evolution of maximum shear rate at the flow wall, $\dot{\gamma}_w$ as a function of flow rate $Q$ (shaded region achievable in the passive limit) for different viscosities of ferrofluids $(\eta_f = 0.01\ \text{Pa}\cdot\text{s to } 10\ \text{Pa}\cdot\text{s})$. For $n = 0.5$ (Fig. 4a)) and $n = 2$ (Fig 4(b)), the $\dot{\gamma}_w$ can be tuned by more than two orders of magnitude. Increasing $\eta_f$ transfers the viscous shear in the transported non-Newtonian material, such that at $\eta_f = \infty$ (red line), the viscous dissipation is entirely in the transported material (analogous to a rigid wall).

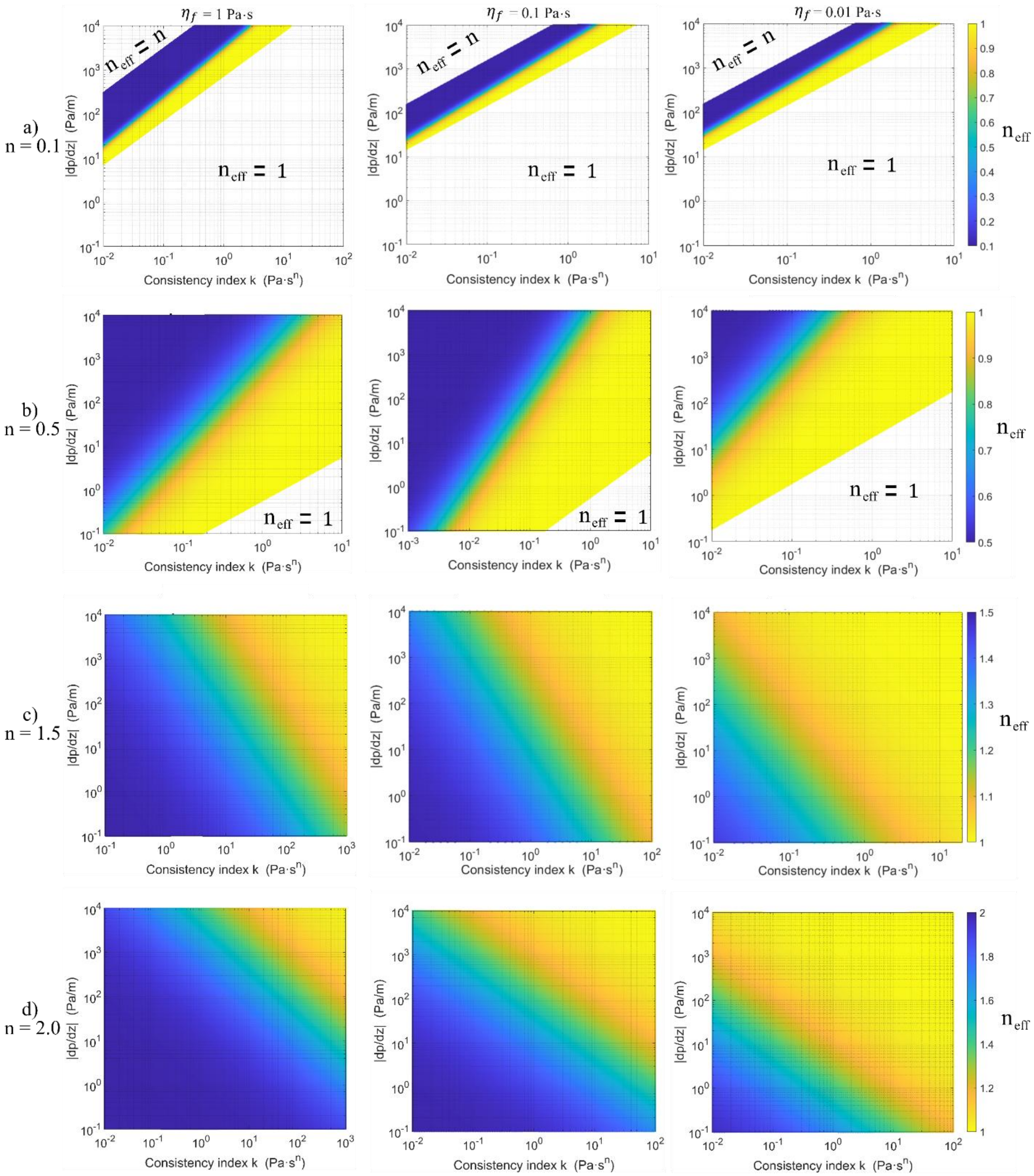

Figure 5 Continuous renormalization of non-linear materials. Effective power law index $n_{eff}$ as a function consistency index $k$, pressure gradient $\left|\partial P/_{\partial z}\right|$, material intrinsic power law index $n$ and lubricant viscosity $\eta_f$. Columns are for three different $\eta_f$= 0.01 Pa · s, 0.1 Pa · s and 1 Pa · s. Rows are for a) $= 0.1$ , b) $n = 0.5$ , c) $n = 1.5$ and d) $n = 2$. Geometric parameters remain the same. Diameter of the core region and cavity remains $d = 1$ mm and $D_c = 1.5$ mm respectively. Pressure gradient remains below gravitational.

Beyond tuning $\dot{\gamma}_w$, an important observation is the non-linear slope of the flow curve, $\left|\partial P/\partial z\right|$ versus $Q$. The slope in the log-log plot gives the flow behavior index ($n$) for a power law material transport (see SI S3 for details). Note that this slope is constant for a rigid wall system (red line), because the flow behavior index $n$ is a material property that does not change with flow rate. However, for lubricated flow, the slope ($n_{eff}$) evolves with $\eta_f$ and $Q$. For $\eta_f = 0.01 \text{ Pa} \cdot \text{s}$, and a large pressure gradient with in the passive limit, the slope of the flow curve $n_{eff}$ approaches → 1. This indicates an effective re-normalization of the material behavior toward a Newtonian regime for both $n = 0.5$ and $n = 2$, where the intrinsic nonlinearity is gradually reduced, as captured by the evolution of $n_{eff}$ toward unity. The $n_{eff}$ hence depends on the material rheology $(n, k)$, flow parameter $(Q, \partial P/\partial z)$ and boundary lubrication $(\eta_f)$. Fig. 5 shows the variation of $n_{eff}$ for shear-thickening and shear-thinning materials with different material characteristics $(n, k, \eta_f)$ and for applied pressure gradient below the passive limit ≤ 10000 Pa/m. It shows that for the present geometric arrangements, in the passive limit, $n_{eff}$ converges to $\approx 1$, indicating a gradual shift from nonlinear to more linear ( Newtonian-like) flow behavior. The material can be apparently tuned to behave like Newtonian $(n_{eff} \approx 1)$ by adjusting the pressure gradient, and $\eta_f$ for a fixed bulk material characteristic $(n, k)$. For shear-thinning materials shown in Fig. 5 a) and Fig. 5b) $(n < 1)$, the flow approaches Newtonian behavior $(n_{eff} \approx 1)$ at large $k$ and high $\left|\partial P/\partial z\right|$, whereas for a shear-thickening material (Fig. 5 c) and (Fig. 5 d), larger k and low $\left|\partial P/\partial z\right|$. Note that our upper limit for pressure gradient is the gravitational limit; a lower pressure gradient can be achieved by simply tilting the flow channel with respect to the vertical. The thickness of regimes of $n_{eff}$ also evolves with $\eta_f$ (compare columns in Fig. 5) as it governs the shear in the non-Newtonian material.

**Yield stress materials:** The gravity-driven transport of yield stress materials with $\tau_y = 1.5$ Pa and $k = 2$ Pa· $s^n$ is presented in Fig. 6 using computational fluid mechanics simulations (CFD, ANSYS CFX) and analytical solutions. Fig. 6a) and Fig.6b) show non-lubricated and lubricated flows respectively. The geometric and forcing parameters remains the same. However, the resulting flow rate is different due to finite yield stress. Flow rate through non-lubricated arrangement (Fig. 6 a)) is $Q_r = 0.64$ $\mu$l/min and 3.18 $\mu$l/min for $n = 0.5$ and $n = 2$ respectively. Due to the no-slip condition, the wall velocity is zero and due to fixed pressure gradient ($-10^4$ Pa/m) wall shear stress is $\tau_w = -\frac{R}{2}\frac{\partial P}{\partial z}$ = 2.5 Pa. Since, $\tau_w > \tau_y$, the material has yielded near the boundary (finite velocity gradient) and is in a jammed state close to the center of the channel, in plug flow regime. The behavior is qualitatively similar for shear-thickening and shear-thinning materials; differences appear in the magnitude of the velocity in the yielded part, and consequently, the plug velocity. The two velocity profiles are distinct, showing signatures of non-linear material property. The solution of the Stokes equation gives the classical prediction for this flow (Eq. 16, Eq. 19 and Eq. 20).

For lubricated arrangement, the distinction is difficult. The velocity profiles for $n = 0.5$ and $n = 2$ are nearly identical, unlike the rigid tube. This shows that the intrinsic non-linearity of the material is suppressed due to lubrication. The Stokes solution (Eq. 16, Eq. 19, Eq. 20) agrees with the CFD solution in the core region with non-Newtonian material and in the annulus with lubricant, for both shear-thinning and shear-thickening materials. Another remarkable observation is the enhancement in flow rate for lubricated flow, $Q = 729.69$ $\mu$l/min and 732.23 $\mu$l/min for $n = 0.5$ and $n = 2$, respectively, which is almost 3 orders of magnitude higher than the rigid wall counterpart ($Q_r$). Note that a similar flow rate in a rigid wall confinement would require orders of magnitude higher pressure gradient $\left|{}^{\partial P}/_{\partial z}\right| \approx 10^8$ Pa/m, which would yield the entire material, even away from the flow wall. Hence, with the lubricated arrangement we can effectively modify the extent of yielded region of the materials. Since $\tau_w = -\frac{R}{2}\frac{\partial P}{\partial z}$, and a large flow rate is still possible with a fraction of $\frac{\partial P}{\partial z}$, the wall stress can be tuned $(\tau_y < \tau_w < \tau_y)$ while maintaining order of magnitude higher throughput.

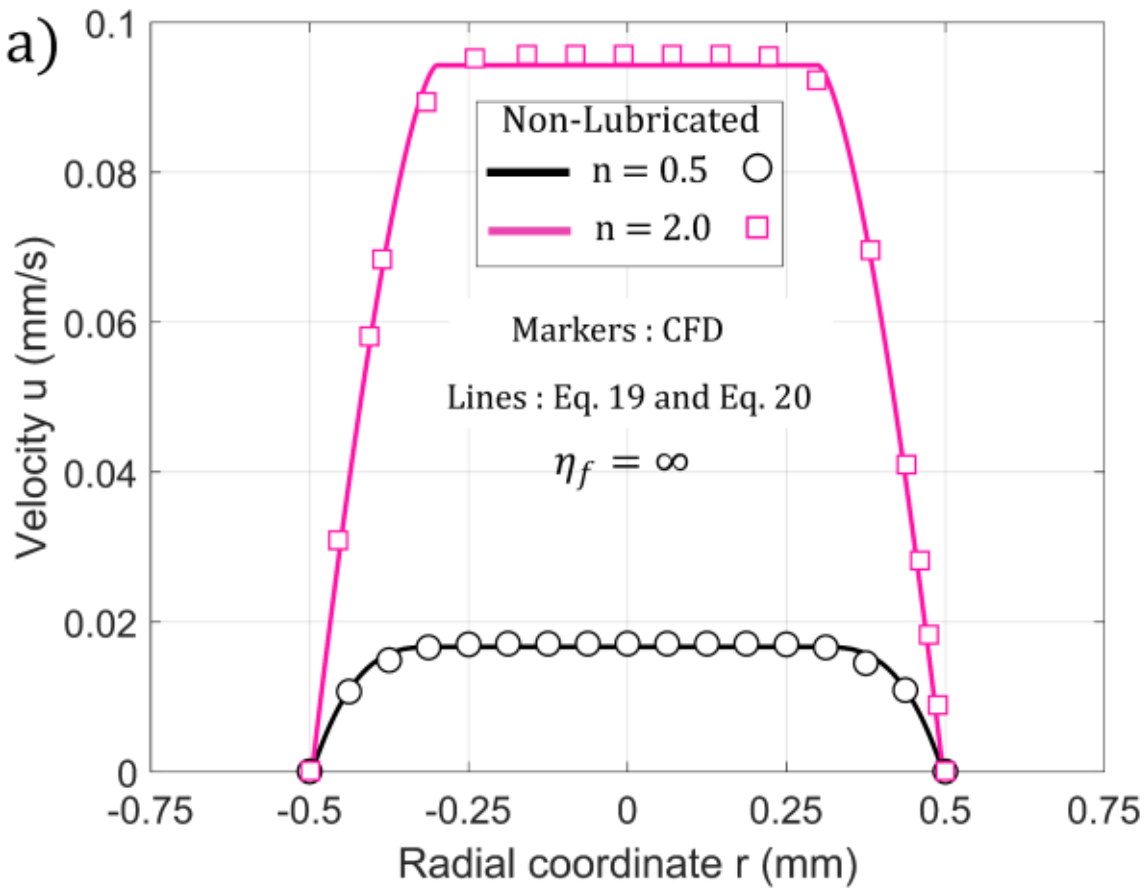

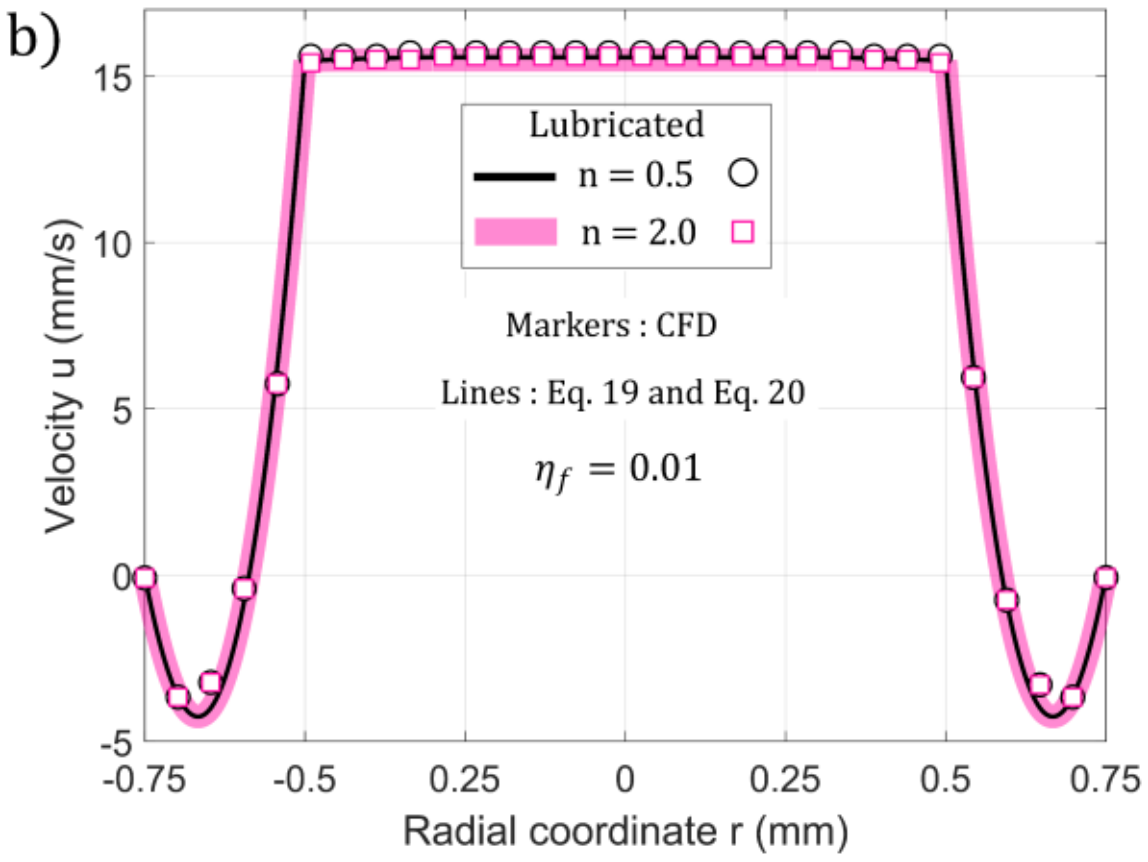

Figure 6 Flow of yield stress material, CFD simulation and Stokes solution (Eq. 16, Eq. 19 and Eq. 20). a) and b) shows velocity field in a fully developed flow through a non-lubricated and lubricated arrangement respectively. The core diameter D, consistency index k and pressure gradient are fixed at 1 mm, 2 Pa$\cdot$ s$^{\mathrm{n}}$ and $-10^4$ Pa/m. Non-Newtonian materials are with $\mathrm{n} = 0.5$ and $\mathrm{n} = 2$. Lines are analytical and markers are CFD simulations. Thick line for $\mathrm{n} = 2$ is to highlight the overlapping in lubricated case.

**Time-dependent Yielding:**

The flow of time-dependent rheological materials (thixotropic) in a confined geometry using CFD is shown in Fig. 7. The simulation is unsteady, and the details are provided in SI S1. In a rigid wall tube of length $L$ and diameter $D$, the flow rate $Q_{th}$ is set to be 7.36 $\mu L/min$, with $\eta_0 = 5$ Pa$\cdot$ s and $\eta_\infty = 0.5$ Pa$\cdot$ s as viscosities at zero shear and maximum shear, respectively. In case of time-dependent rheological materials, the response depends on the residence time $\alpha = \frac{L}{U}$ of the material in the confined channel, where $U$ is the average velocity of flow. The Deborah number ($De$) becomes important, defined as $\alpha/\tau_{\text{struct}}$, is the ratio of the time it takes for a material to adjust to applied stresses or deformations in the flow channel.

In case of the non-lubricated arrangement (Fig. 7a), $\alpha$ is 64 s, which is greater than the $\tau_{\text{struct}}$ with $De = 12.8$. This results in shear induced breaking of the structure in the material, quantified by the structure parameter $S$. Due to high shear near the flow wall, $S$ approaches 0, whereas away from the wall S increases towards 1 (Fig. 7a)). The structure factor evolves over the length of flow linked to residence time (compare panels in Fig. 7a)). The structural relaxation time is significantly smaller than the residence time ($De \gg 1$), leading to rapid structural equilibration with $S \ll 1$ corresponding to broken structure within a small fraction of the to pressure gradient with residence time, a and c are the pressure gradient values at minimum and maximum residence time respectively. a = 19387 Pa/m, c = 6154 Pa/m . $t_c = 0.81\ s$ is the characteristic decay time.

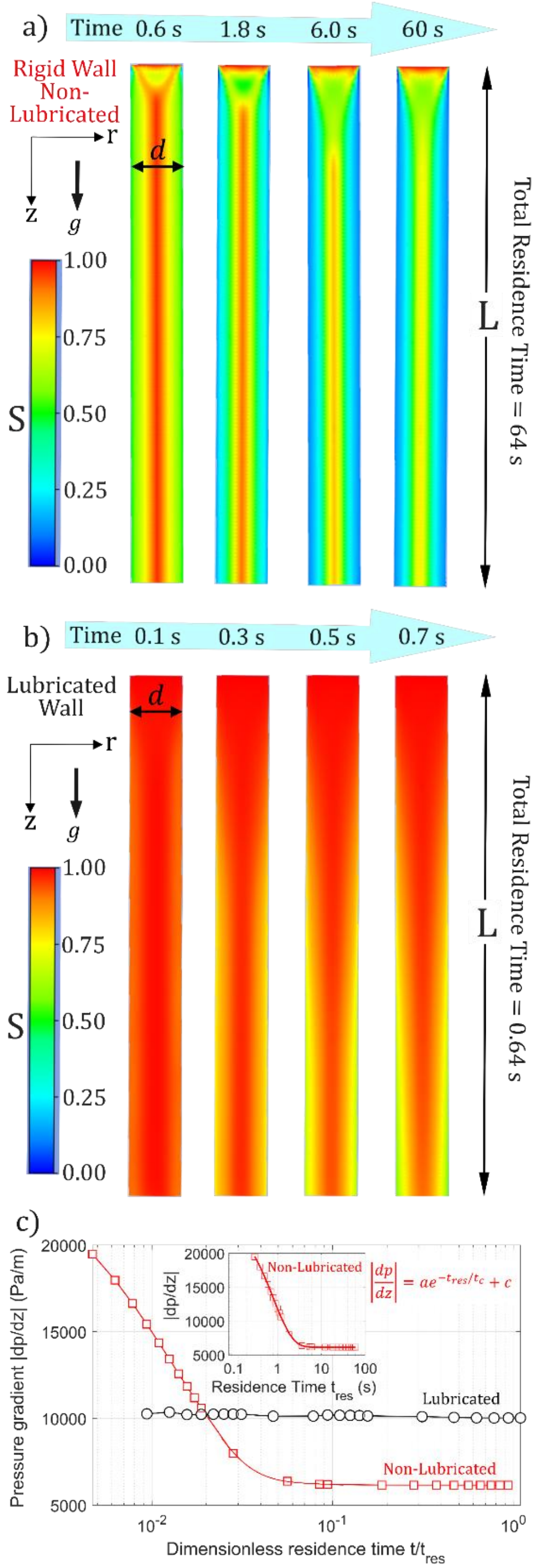

Figure 7. Temporal evolution of thixotropic material transport. Structure parameter $S$ in a) non-lubricated, b) Lubricated flows. c) Evolution of pressure gradient over dimensionless residence time for non-lubricated and lubricated flows. Inset shows the exponential fit

channel length. In contrast, for lubricated flow (Fig. 7b), with $\eta_f = 0.01\ \mathrm{Pa \cdot s}$, the residence time is two orders of magnitude lower due to an increase in the flow rate $Q = 718\ \mu L/min$. The residence time is 0.64 s ($De \ll 1$), suppressing structural evolution along the flow. Furthermore, since the shear is confined in the ferrofluid layer, we do not observe a significant change in the $S$ over time, and material flows in a jammed state with almost intact structure. $S$ remains close to 1, over time and along the flow channel. Thus, despite the inherently time-dependent rheology, the flow effectively exhibits time-independent behavior, with the material remaining in a jammed state even at high flow rates. Moreover, since the viscosity is linked to the structure, it evolves over time (along the flow length $L$), hence the pressure gradient required to maintain a constant $Q_{th}$ (non-lubricated) or $Q$ (lubricated) must also be a function of time. This needs real-time feedback to maintain steady flow rate. Figure 7 c) shows the evolution of the pressure gradient over time for a constant $Q_{th}$ or $Q$. For a non-lubricated, the pressure gradient reduces over time with a characteristic decay time $t_c = 0.81\ s$ (see inset) that is linked to the time scale of equilibration $1/(k_b + k_d\dot{\gamma})$. For lubricated wall, the pressure gradient is almost constant over time. Due to smaller shear and intact structure, no active feedback is required to maintain a constant $Q$. Inset shows the exponential fit expected for non-lubricated flows. Note that we have chosen $Q_{th}$ and $Q$ such that the pressure gradient remains of the order of gravitational pressure gradient.

**Discussion:** In this study, numerical modelling and analytical solutions of the Stokes equation are presented for the flow of time-independent and time-dependent non-Newtonian fluids in cylindrical confinement, with a particular focus on lubrication via a colloidal magnetic liquid (ferrofluid) annulus.

For pure power-law fluids with zero yield stress, the study demonstrates that lubrication enables up to two orders of magnitude increase in flow rate under the same pressure gradient. Moreover, lubricated systems exhibit a transition toward Newtonian-like behavior as lubricant viscosity decreases, regardless of the intrinsic nonlinearity of the core material. This “rheological renormalization” depends not only on the ferrofluid viscosity, but also on the applied pressure gradient and the bulk material parameters $(n, k)$.

For yield-stress fluids, gravity-driven flow in non-lubricated rigid-walled tubes results in yielding of the material near the wall, while the central region remains unyielded, forming a plug flow whose magnitude strongly depends on the rheological index $n$. Introducing a magnetically confined ferrofluid annulus creates a lubricated, boundary that is mechanically stable. This lubrication increases flow rates by approximately three orders of magnitude under the same pressure gradient and suppresses differences between shear-thinning and shear-thickening behavior, resulting in nearly identical velocity profiles.

For time-dependent (thixotropic) materials, a structural kinetics model is incorporated via a transport equation for a structure parameter $S$. In rigid channels, high shear leads to rapid structural breakdown (low $S$), time-dependent viscosity and requiring dynamic pressure control to maintain constant flow conditions. In contrast, lubricated flow significantly reduces shear and residence time, suppressing structural evolution (high $S$) and yielding effectively time-independent behavior without the need for active feedback.

It is also shown that parameter regimes exist in which magnetic lubrication decouples the apparent flow performance from the intrinsic non-Newtonian properties of the fluid. This enables passive, high-throughput transport with minimal energy input, with potential applications in microfluidics, soft material handling, and transport of complex fluids where conventional pumping is inefficient or impractical.